\documentclass[showpacs,twocolumn,preprintnumbers,amsmath,amssymb,prr ]{revtex4-2}

\usepackage{graphicx}
\usepackage{nicefrac}
\usepackage{color}
\usepackage{braket}
\usepackage[colorlinks=true,urlcolor=blue,citecolor=blue,linkcolor=blue]{hyperref}

\usepackage{booktabs}

\begin{document}

\title{
Bipartite entanglement extracted from multimode squeezed light generated in lossy waveguides
}

\author{Denis A. Kopylov}
\email{denis.kopylov@uni-paderborn.de}
\affiliation{Department of Physics, Paderborn University, Warburger Str. 100, D-33098 Paderborn, Germany}
\affiliation{Institute for Photonic Quantum Systems (PhoQS), Paderborn University, Warburger Str. 100, D-33098 Paderborn, Germany} 

\author{Torsten Meier}
\affiliation{Department of Physics, Paderborn University, Warburger Str. 100, D-33098 Paderborn, Germany}
\affiliation{Institute for Photonic Quantum Systems (PhoQS), Paderborn University, Warburger Str. 100, D-33098 Paderborn, Germany} 

\author{Polina R. Sharapova}
\affiliation{Department of Physics, Paderborn University, Warburger Str. 100, D-33098 Paderborn, Germany}

\date{\today}

\begin{abstract}
Entangled two-mode Gaussian states constitute an important building block for continuous variable quantum computing and communication protocols.  
In this work, we theoretically study two-mode bipartite states which are extracted from multimode light generated via type-II parametric down-conversion (PDC) in lossy waveguides.
For these states, we demonstrate that the squeezing quantifies entanglement and we construct a measurement basis which results in the maximal bipartite entanglement.
We illustrate our findings by numerically solving the spatial master equation for PDC in a Markovian environment.
The optimal measurement modes are compared with two widely-used broadband bases:
the Mercer-Wolf basis (the first-order coherence basis) and the Williamson-Euler basis.
\end{abstract}

\maketitle

\section{\label{sec_intoduction } Introduction }

Entangled two-mode bipartite Gaussian   states play a fundamental role in quantum optics as they represent building blocks for many quantum optical protocols~\cite{Braunstein_2005_RMP,Weedbrook_2012}.
An efficient source of highly entangled states is thus an essential part for scalable quantum technological applications~\cite{Usenko_2025}. 
Currently, one of the most promising platforms for producing Gaussian quantum light is based on single-path pulsed parametric down-conversion (PDC) in nonlinear waveguides with second-order nonlinear susceptibility~\cite{Stokowski_2023,Henry2023,Roeder_2024,Placke_2024,Schuhmann_2024_PRX}.
These platforms potentially allow for the generation of
ultrashort pulsed squeezed vacuum states with a high-repetition rate.
However, two main factors are limiting the performance of such sources.
The first one is the presence of scattering losses,
i.e., the intrinsic waveguide losses~\cite{Hammer_2024, Melati_2014},
which are caused by technological imperfections during the fabrication and result in the generation of a mixed state in the PDC process.
The second limiting factor is the multimode structure of the generated PDC light~\cite{Christ_2013,Sharapova_2015,Horoshko_2019,Kopylov_2025,Kopylov_2025_PRR}, which is induced by nonlinear optical coupling of interacting fields and strongly depends on the waveguide dispersion and the pump profile.
To accurately operate with multimode PDC light, it is necessary to determine the correct mode shapes, since measurements in the wrong modes can lead to the loss of non-classicality~\cite{Christ_PRA_2014,Kopylov_2025,Kala_OE_2025}.
One of the tasks where mode shapes play a crucial role is the generation of the two-mode bipartite states (TMBS) based on multimode type-II PDC light.
Experimentally, TMBS can be obtained from the initial multimode bipartite light by using a single quantum pulse gate (QPG)~\cite{Brecht_2015,Reddy_2018,Serino_2023} for each partition.
Alternatively, one can selectively work with the chosen two modes of the multimode light by implementing proper gates with the use of homodyne detection~\cite{Kouadou_2023,Roh_2025}.

In this paper, we study two-mode bipartite Gaussian states extracted from pulsed type-II multimode PDC generated in lossy waveguides.
We show how to choose the measurement basis for obtaining the maximally-entangled two-mode bipartite state (TMBS), which then can be used for a proper homodyne detection or QPG protocol.

\section{\label{sec:theory}Theoretical description}

\paragraph*{Theoretical model for type-II PDC.} For the description of PDC generated in lossy media, we use the numerical scheme developed in Refs.~\cite{Kopylov_2025, Kopylov_2025_PRR}. 
The details of this approach are given in the Appendices~\ref{appendix_a} and \ref{appendix_b}.
In the framework of multimode Gaussian states, the type-II PDC is described as a multimode bipartite system: part A (also called as signal) and part B (also called as idler), each consisting of $N$ bosonic monochromatic modes,  $\hat{\mathbf{a}} = \big(\hat{a}_0, \hat{a}_1, \dots, \hat{a}_N\big)^T$ and $\hat{\mathbf{b}} = \big(\hat{b}_0, \hat{b}_1, \dots, \hat{b}_N\big)^T$, respectively.
The lower indices correspond to the different discrete frequencies.
The quantum state generated via PDC from the vacuum state is fully described by the correlation matrices $\mathcal{D}$ and $\mathcal{C}$, which obey the spatial master equations. 
For a type-II PDC process in a Markovian environment, see Ref.~\onlinecite{Kopylov_2025_PRR} and Appendix~\ref{appendix_b}, their solution has the form
\begin{equation}
    \mathcal{D} =     
    \begin{pmatrix}
        \braket{\mathbf{\hat{a}^\dagger \hat{a}} }  & \mathbf{0}  \\
        \mathbf{0}  & \braket{\mathbf{\hat{b}^\dagger \hat{b}} } 
    \end{pmatrix}, ~ 
    \mathcal{C} =     
    \begin{pmatrix}
        \mathbf{0}  & \braket{\mathbf{\hat{a} \hat{b}} }  \\
        \braket{\mathbf{\hat{b} \hat{a}} }  & \mathbf{0} 
    \end{pmatrix},
    \label{eq_solution_block}
\end{equation}
where the matrices $\braket{\mathbf{\hat{a}^\dagger \hat{a}} }$, $\braket{\mathbf{\hat{b}^\dagger \hat{b}} }$ and $\braket{\mathbf{\hat{a} \hat{b}} }$ are the correlation matrices with elements $\braket{{\hat{a}^\dagger_i \hat{a}}_j }$, $\braket{{\hat{b}^\dagger_i \hat{b}_j} }$ and $\braket{{\hat{a}_i \hat{b}_j} }$, respectively.
The expectation values $\braket{\mathbf{\hat{a}}^\dagger}=\braket{\mathbf{\hat{a}}}=\braket{\mathbf{\hat{b}}^\dagger}=\braket{\mathbf{\hat{b}}}\equiv0$ due to the initial vacuum state.
Note that the presence of zero matrices $\mathbf{0}$ in Eq.~\eqref{eq_solution_block} is a property of the considered type-II PDC. 
Indeed, in this case, the PDC Hamiltonian contains only the interaction terms between the signal and idler subsystems.
Below we restrict ourselves to type-II PDC and therefore use Eq.~\eqref{eq_solution_block} as the starting point for further derivations. 

Note that in this study, we only focus on losses during PDC generation; however, the linear external losses, e.g., transmission and detection losses, could easily be included in the consideration.
The procedure for introducing frequency-dependent external losses is described in detail in Ref.~\onlinecite{Kopylov_2025}.
Here, we only mention that the external losses are included via the transformations $\hat{a}_n \rightarrow t^{a}_n\hat{a}_n$ and $\hat{b}_n \rightarrow t^{b}_n\hat{b}_n$, where $t^{a}_n$ and $
t^{b}_n$ are the complex field transmission coefficients.
These transformations do not change the structure of Eq.~\eqref{eq_solution_block}; therefore, all the results we obtained are also valid for frequency-dependent external losses.
 
Alternatively to the matrices $\mathcal{D}$ and $\mathcal{C}$, the generated PDC state can be presented in terms of the covariance matrix $\Sigma$.
The relationship between the matrices $\mathcal{D}$, $\mathcal{C}$, and  $\Sigma$ are given in Appendix~\ref{appendix_a}.

\begin{figure}
    \includegraphics[width=0.99\linewidth]{ 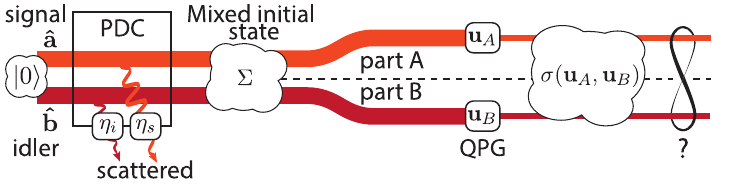 } 
    \caption{The scheme of a TMBS preparation from a multimode type-II PDC process. After the lossy PDC process, a multimode mixed state with the covariance matrix $\Sigma$ is generated. 
    Two quantum pulse gates (QPG) in partition A with the mode $\mathbf{u}_A$ and in partition B with the mode $\mathbf{u}_B$ provide the TMBS with the covariance matrix $\sigma(\mathbf{u}_A, \mathbf{u}_B)$.
        }
    \label{fig_1}
\end{figure}

\paragraph*{Two-mode bipartite Gaussian state.}

Let us assume that we measure the light in only two modes: one mode in each of subsystems A and B.
This scenario can experimentally be realized using the QPGs as shown in Fig.~\ref{fig_1} or in homodyne detection, where the modes of local oscillators correspond to measurement modes. 
As a result, instead of using the initial multimode state, we effectively work with a TMBS built by two measurement modes.
In the general case, these modes are represented as a superposition of the initial monochromatic PDC modes, therefore, mathematically they can be found using a passive local transformation 
\begin{equation}
     U = \begin{pmatrix}
\mathbf{u}_A & \mathbf{0}_{N}  \\
            \mathbf{0}_{N}  & \mathbf{u}_B
\end{pmatrix} ,
\end{equation}
where $\mathbf{u}_A$ and $\mathbf{u}_B$ are the normalized row-vectors of size $N$. 
Then the annihilation operators for the broadband measurement modes are given by $\hat{A} = \mathbf{u}_A \hat{\mathbf{a}}$ and $\hat{B} = \mathbf{u}_B \hat{\mathbf{b}}$.

Applying the transformation $U$ to the initial multimode system Eq.~\eqref{eq_solution_block}, via the rules~\eqref{eq_general_transformation}, one obtains the following nonzero second-order correlators for the introduced broadband modes:
\begin{align}
    \label{eq_single_transformations_1}
    \braket{\hat{A}^\dagger \hat{A} } & = \mathbf{u}_A^*  \braket{\mathbf{\hat{a}^\dagger \hat{a}} } \mathbf{u}_A^T, \\
    \braket{\hat{B}^\dagger \hat{B} } & = \mathbf{u}_B^*  \braket{\mathbf{\hat{b}^\dagger \hat{b}} } \mathbf{u}_B^T, \\
    \braket{\hat{A} \hat{B} } & = \mathbf{u}_A \braket{\mathbf{\hat{a} \hat{b}} } \mathbf{u}_B^T. 
    \label{eq_single_transformations_3}
\end{align}
To simplify the following expressions, we choose the phases of the modes in such a way that $\textrm{Im}(\braket{\hat{A}\hat{B}})=0$.
This can be achieved via additional phase rotations $\mathbf{u}_A \rightarrow \mathbf{u}_A e^{i\phi_A}$ and $\mathbf{u}_B \rightarrow \mathbf{u}_B e^{i\phi_B}$.
The number of photons in the signal and idler parts are $N_A = \braket{\hat{A}^\dagger \hat{A} }$ and $N_B = \braket{\hat{B}^\dagger \hat{B} }$, respectively.
Note that the vectors $\mathbf{u}_A$ and $\mathbf{u}_B$ fully determine the TMBS: choosing different measurement modes, we obtain different TMBS.

By introducing the quadrature operators for broadband modes $\hat{q}^A = \hat{A}^\dagger + \hat{A} $, $\hat{p}^A =  i(\hat{A}^\dagger - \hat{A}) $ and $\hat{q}^B = \hat{B}^\dagger + \hat{B} $, $\hat{p}^B =  i(\hat{B}^\dagger - \hat{B}) $  (we use $\hbar=2$, the commutation relations $[\hat{q}^A, \hat{p}^A] = [\hat{q}^B, \hat{p}^B] = 2i$, $[\hat{q}^A, \hat{p}^B] = 0$), the covariance matrix for the corresponding TMBS reads
\begin{equation}
    \sigma(\mathbf{u}_A, \mathbf{u}_B) = 
        \begin{pmatrix}
            \alpha  &  0  &  \gamma  &  0    \\
            0  &  \alpha  &  0  & -\gamma  \\
            \gamma  &  0  &  \beta  &  0    \\
            0  & -\gamma  &  0  &  \beta    \\
        \end{pmatrix}, 
        \label{eq_cov_matrix_1x1}
\end{equation} 
where we order the quadratures as $ (\hat{q}^A, \hat{p}^A, \hat{q}^B, \hat{p}^B) $. 
The parameters of the matrix are defined as follows $ \alpha\equiv \braket{\hat{p}^A \hat{p}^A} = \braket{\hat{q}^A \hat{q}^A} = 1 + 2\braket{\hat{A}^\dagger \hat{A} } $, $\beta \equiv \braket{\hat{p}^B \hat{p}^B} = \braket{\hat{q}^B \hat{q}^B}  = 1 + 2\braket{\hat{B}^\dagger \hat{B} }$ and $\gamma \equiv  \braket{\hat{q}^A \hat{q}^B} = -\braket{\hat{p}^A \hat{p}^B}=2 \braket{\hat{A} \hat{B} }$.

Note that the condition $\textrm{Im}(\braket{\hat{A}\hat{B}})=0$ leads to the absence of correlations between the coordinates and momenta of different subsystems, namely $\braket{\hat{q}^A \hat{p}^B}=\braket{\hat{p}^A \hat{q}^B}=0$, see Appendix~\ref{appendix_a}.
Moreover, the covariance matrix Eq.~\eqref{eq_cov_matrix_1x1} represents a special case of the so-called standard form covariance matrix~\cite{Simon_2000,Duan_2000}, which significantly simplifies the characterization of entanglement.

\paragraph*{Entanglement in TMBS.}

Since different broadband modes provide different TMBS, the problem of extracting the TMBS with the maximal degree of entanglement $E_M$ reduces to maximizing a proper entanglement measure~\cite{Adesso_2005,Plenio_2007} $\mathcal{E}\left(\sigma\right)$ by varying the local transformation $U$ 
\begin{equation}
    E_M = \max_{\substack{\mathbf{u}_A, \mathbf{u}_B: \\ |\mathbf{u}_A|=1, |\mathbf{u}_B|=1} } \mathcal{E}\left(\sigma(\mathbf{u}_A, \mathbf{u}_B)\right).
    \label{eq_optimization}
\end{equation}  
In this work, as the entanglement measure, we use the logarithmic negativity~\cite{Vidal_2002, Plenio_2005_LG} 
\begin{equation}
    \mathcal{E}(\sigma) = \max\left[-\log\left(\nu_-(\tilde{\sigma})\right), 0 \right],
\end{equation}
where the value $\nu_-(\tilde{\sigma})$ is the minimal symplectic value of a partially transposed covariance matrix $\tilde{\sigma} = T \sigma T$ with  $T=\mathrm{diag}(1,1) \oplus \mathrm{diag}(1,-1)$.
The symplectic spectrum of the partially transformed covariance matrix ${\nu(\tilde\sigma})$ corresponds to the eigenspectrum of the matrix~\cite{Vidal_2002, Adesso_2005,Weedbrook_2012} $|i \Omega \tilde{\sigma}|$, where $\Omega = \omega \oplus \omega$ and
$\omega=\big(\begin{smallmatrix}
  0 & 1\\
  -1 & 0
\end{smallmatrix}\big)$.
For the matrix $\sigma(\mathbf{u}_A, \mathbf{u}_B)$, the corresponding characteristic polynomial  reads  
$p(x) = x^4 - (\alpha^2 + \beta^2 + 2\gamma^2) x^2  + (\alpha \beta - \gamma^2)^2 = 0$ and results in the symplectic values  
\begin{equation}
    \nu_\pm(\tilde{\sigma})=\frac{1}{2} \left(\alpha+\beta\pm\sqrt{(\alpha-\beta)^2+4 \gamma^2}\right).
    \label{symplectic}
\end{equation}

\paragraph*{Squeezing in TMBS.}
The eigenvalues $\lambda(\sigma)$ of the covariance matrix $\sigma$ are determined by the roots of its characteristic polynomial, which for the matrix Eq.~\eqref{eq_cov_matrix_1x1} reads 
$p(\lambda) = \det(\sigma-\lambda \mathbf{1}) = \left((\lambda-\alpha)(\lambda-\beta) -\gamma^2\right)^2 = 0$.
The smallest eigenvalue $\lambda_{-}(\sigma)$ determines the maximal degree of squeezing~\cite{Simon_1994, Wolf_2003} contained in the system. 
The states with $\lambda_{-}(\sigma)<1$ are called the squeezed states  (in our units, the variance of the vacuum state equals to 1).
Note that a higher degree of squeezing corresponds to a lower eigenvalue (lower quadrature variance).
In what follows, the squeezing is given in dB as $10 \log_{10}(\lambda_{-})$.

\paragraph*{Connection between squeezing and entanglement for type-II PDC.}

Comparing the eigenvalues of the covariance matrix Eq.~\eqref{eq_cov_matrix_1x1} with the symplectic spectrum of the partially transformed covariance matrix Eq.~\eqref{symplectic}, we notice that they coincide, namely, 
\begin{equation}
  \lambda_{\pm}(\sigma) \equiv \nu_\pm(\tilde{\sigma}).
  \label{eq_connection}  
\end{equation}
As a result, the maximal degree of squeezing (minimal eigenvalue $\lambda_{-}(\sigma)$) is a proper entanglement measure for the studied TMBS.
This means that the optimization task Eq.~\eqref{eq_optimization} can be reformulated in terms of finding the maximal squeezing of the initial multimode state.

\paragraph*{Connection between squeezing and entanglement for arbitrary system.}

In general, the connection between squeezing and entanglement is not as trivial and was studied, e.g., in Ref.~\onlinecite{Wolf_2003}:
Squeezing is present in entangled Gaussian states, while the opposite is not always true.
As an example, we consider the TMBS consisting of two single-mode systems with the covariance matrix
$\sigma_{sm}=\big(\begin{smallmatrix}
  e^{r} & 0\\
  0 & e^{-r}
\end{smallmatrix}\big) 
    \oplus 
\big(\begin{smallmatrix}
  e^{r} & 0\\
  0 & e^{-r}
\end{smallmatrix}\big)  $.
This state is squeezed but is not entangled. 
Therefore, the presence of single-mode squeezing in one of the subsystems breaks Eq.~\eqref{eq_connection}.  

To answer the question why Eq.~\eqref{eq_connection} holds for our system, we notice that the initial multimode system Eq.~\eqref{eq_solution_block} does not contain single-mode squeezing in either part A or B.
Indeed, the correlation matrices $\braket{\mathbf{\hat{a} \hat{a}} } = \braket{\mathbf{\hat{b} \hat{b}} } =  \mathbf{0} $, which physically comes from the absence of signal-signal (and idler-idler) terms in the type-II PDC interaction Hamiltonian and resulting master equations.
As a result, the inequalities $\braket{(\Delta \hat{p}^A)^2}=\braket{(\Delta \hat{q}^A)^2}\geq1$ and $\braket{(\Delta \hat{p}^B)^2}=\braket{(\Delta \hat{q}^B)^2}\geq1$ hold for any choice of modes $\mathbf{u}_A$ and $\mathbf{u}_B$, indicating the absence of single-mode squeezing.  
Therefore, the only possible squeezing $\braket{(\Delta \hat{p}^F)^2}<1 $ can only be present in some joint mode $\hat{F} = t \hat{A} + r \hat{B}$ with $|t|^2 + |r|^2 = 1$ and $t,r\neq 0$.
This type of squeezing is known as two-mode squeezing.
In turn, the mode $\hat{F}$ can always be chosen to get $\braket{(\Delta \hat{p}^F)^2}=\lambda_{-}(\sigma)$.

\paragraph*{Maximally-squeezed basis.} 
To find the maximally entangled TMBS which can be obtained from the initial multimode system Eq.~\eqref{eq_solution_block}, we have to find the modes with maximal squeezing (minimal eigenvalue) for initial multimode system.
Indeed, the lowest eigenvalue $\lambda_{min}$ of the covariance matrix $\Sigma$ determines the maximal squeezing present in the initial multimode system.
Then, by determining the vectors $\mathbf{u}_A$ and $\mathbf{u}_B$, for which $\lambda_-(\sigma(\mathbf{u}_A,  \mathbf{u}_B)) = \lambda_{min}$, one finds the required shapes of the modes that maximize Eq.~\eqref{eq_optimization}.

The basis that provides maximal squeezing of lossy systems and brings the covariance matrix to the second canonical form~\cite{Simon_1994} is the maximally-squeezed basis (MSq-basis)~\cite{Kopylov_2025}. 
The mode $\hat{F}$ of this basis is found from the eigenvector $\mathbf{v}_{min} = (x^{a}_1, y^{a}_1, \dots, x^{a}_N, y^{a}_N; x^{b}_1, y^{b}_1, \dots, x^{b}_N, y^{b}_N )$ that corresponds to the eigenvalue $\lambda_{min}$, namely $\hat{F} = \sum_n \big(v^a_n \hat{a}_n + v^b_n \hat{b}_n \big)$,
where we have introduced two complex vectors $[\mathbf{v}^a]_n = y^a_n+i x^a_n$ and $[\mathbf{v}^b]_n=y^b_n+i x^b_n$.
Then, the normalized vectors $\mathbf{u}^{MSq}_A = \frac{\mathbf{v}^a}{|\mathbf{v}^a|} $ and $\mathbf{u}^{MSq}_B =  \frac{\mathbf{v}^b}{|\mathbf{v}^b|} $
give us the desired broadband modes that minimize Eq.~\eqref{eq_optimization}.
The vectors $\mathbf{u}^{MSq}_A$ and $\mathbf{u}^{MSq}_B$ define the state called MSq-TMBS and the corresponding covariance matrix $\sigma^{MSq} \equiv \sigma(\mathbf{u}^{MSq}_A, \mathbf{u}^{MSq}_B)$.
The minimal eigenvalues of matrices $\Sigma$ and $\sigma^{MSq}$ coincide, which guarantees that the generated TMBS is a TMBS with the maximal squeezing and, therefore, with the maximal entanglement.

\paragraph*{Other broadband bases.} 
Studies of multimode properties of non-classical optical states are commonly performed with two other broadband bases: the Mercer-Wolf (MW) basis~\cite{Wolf_1982,Mandel_Wolf_book} (also known as the first-order coherence mode basis~\cite{Fabre_2020}) and the Williamson-Euler (WE) basis~\cite{Weedbrook_2012,Hosaka_2016,Safranek_2018,Gupt2019,Houde_2023,houde_2024}.
Based on the MW and WE decompositions, the vectors $\mathbf{u}^{MW}_A$, $\mathbf{u}^{MW}_B$ and $\mathbf{u}^{WE}_A$, $\mathbf{u}^{WE}_B$ can be obtained. 
These vectors define additional TMBSs (for details see Appendix~\ref{appendix_c}) that are further referred to as MW-TMBS and WE-TMBS with the covariance matrices $\sigma^{MW} \equiv \sigma(\mathbf{u}^{MW}_A, \mathbf{u}^{MW}_B)$ and $\sigma^{WE} \equiv \sigma(\mathbf{u}^{WE}_A, \mathbf{u}^{WE}_B)$, respectively.
 
\paragraph*{Special case: pure multimode state.} 
The special case where the Mercer-Wolf, Williamson-Euler, and MSq bases coincide corresponds to a pure multimode state, i.e., to PDC generated in a lossless waveguide~\cite{Kopylov_2025}.
This basis is known as the Schmidt-mode basis~\cite{Fabre_2020,Raymer_2020}.
The first Schmidt mode is a mode with the highest number of photons and, at the same time, with the smallest eigenvalue of the covariance matrix.
Building the TMBS based on the first Schmidt mode, we obtain a pure state that is characterized by a covariance matrix $\sigma^S$ with parameters $\alpha=\beta$ and $\gamma = \sqrt{1-\alpha^2}$ and corresponds to the ideal two-mode squeezer.

\section{Numerical simulations}

\begin{figure}
    \includegraphics[width=0.49\linewidth]{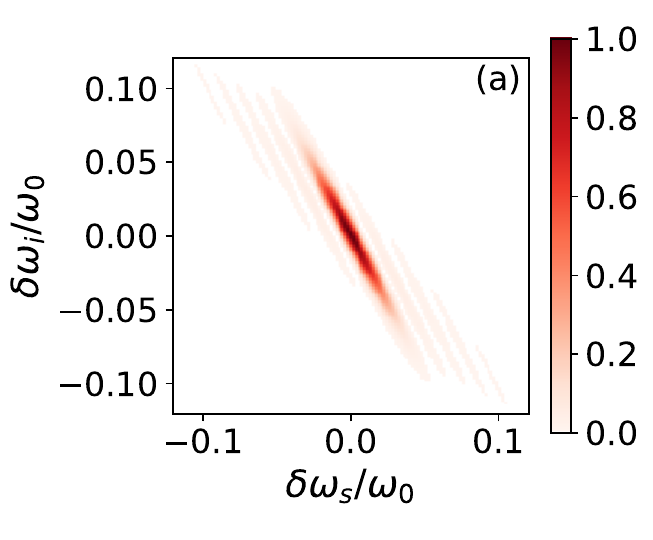 } 
    \includegraphics[width=0.49\linewidth]{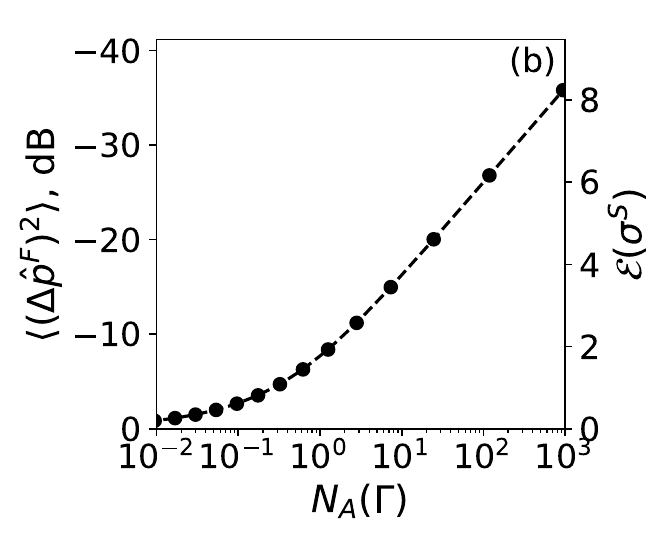 } 
    \caption{ Numerical results for a lossless waveguide WG0, the parameters of which are given in the main text.
        (a) Normalized joint spectral intensity for low-gain PDC, $\delta \omega = \omega - \omega_0$, where $\omega_0=\omega_p/2$ is the central frequency of the PDC light.
        (b) Dependence of the smallest quadrature variance $\braket{(\Delta \hat{p}^F)^2}=\lambda_{-}(\sigma)$ as a function of the number of photons $N_A=N_B$.
        The additional axis shows the corresponding logarithmic negativity $\mathcal{E}(\sigma^S)$.
        }
    \label{fig_2_lossless}
\end{figure}

 \begin{figure}
    \includegraphics[width=0.99\linewidth]{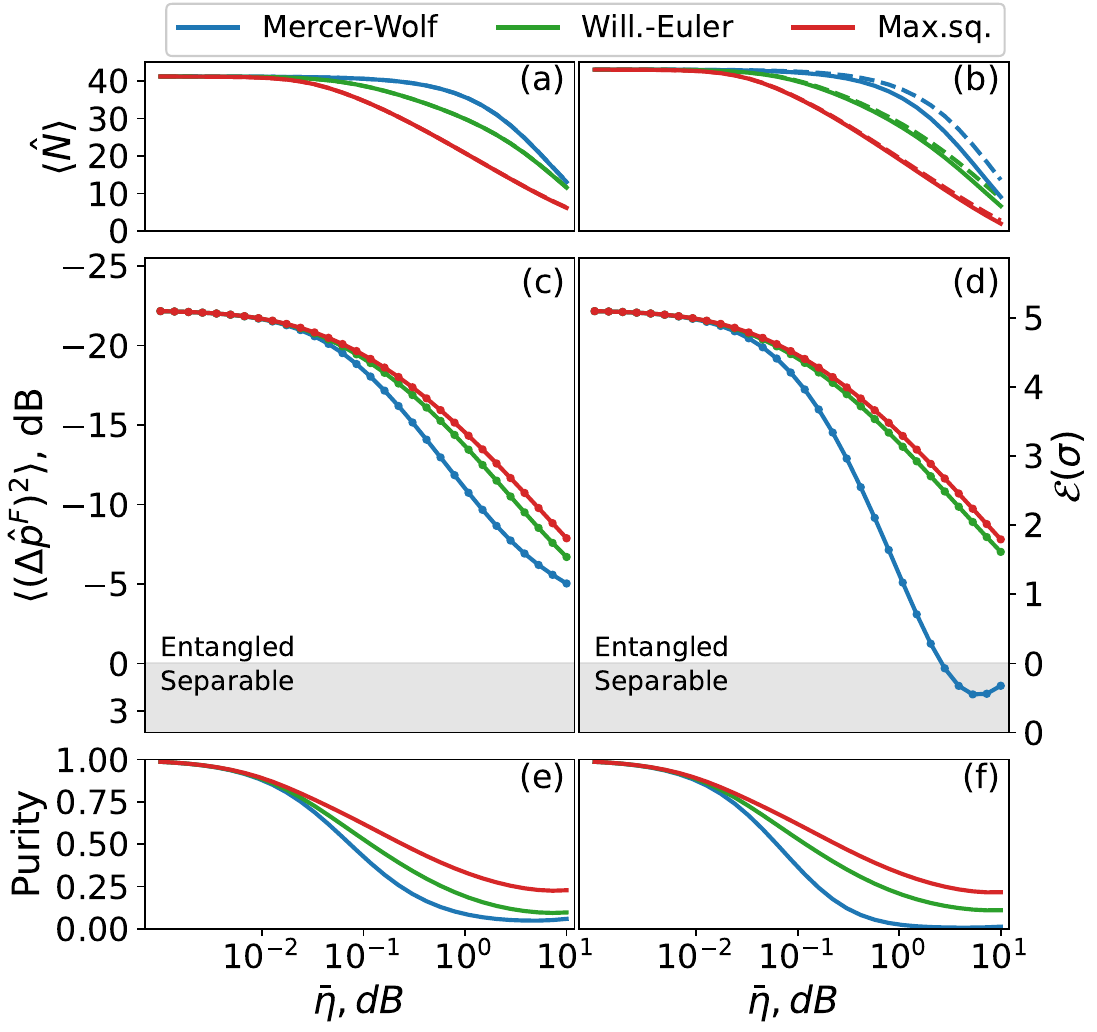 } 
     \caption{
        Numerical results for lossy waveguides: (a,c,e) WG1 and (b,d,f) WG2. The waveguide parameters are given in the main text.
        Different colors correspond to TMBS build with different modes: MW (blue), WE (green), and MSq (red).
        (a, b) The number of photons $N_A$ (solid lines) and $N_B$ (dashed lines) as a function of losses $\bar\eta$.
        (c, d) The dependencies of the smallest quadrature variance $\braket{(\Delta \hat{p}^F)^2} = \lambda_{-}(\sigma)$ for $\sigma^{MW}$, $\sigma^{WE}$ and $\sigma^{MSq}$ as a function of losses $\bar\eta$.
        Additional right axis show the logarithmic negativity $\mathcal{E}(\sigma)$. The gray area highlights the region with zero logarithmic negativity.
        (e, f) Purities of  $\sigma^{MW}$, $\sigma^{WE}$ and $\sigma^{MSq}$ as a function of losses $\bar\eta$.}
 \label{fig_3_alpha_dep}
 \end{figure}

To simulate frequency-degenerate type-II PDC, we consider a $1$~cm long waveguide with manually defined dispersion and losses.
As the pump, we use a Gaussian pulse with a full width at half maximum of $\Delta\tau = 0.4$~ps and a central wavelength of $\lambda_p = 755$~nm.
For simplicity, we limit ourselves to the linear refractive index dependence for each field, which is parametrized as
$
    n(\omega) = n(\omega_0) + \frac{\omega-\omega_0}{\omega_0}\left[\frac{c}{v_g(\omega_0)} - n(\omega_0)\right],
$
where $c$ is the speed of light, $v_g$ is the group velocity, and $\omega_0$ is the central frequency.
For our studies, we choose the following parameters for the pump, signal, and idler fields (indices $p$, $s$ and $i$, respectively): $n_p = n(\omega_p) = 1.9$, $v_g^p = 0.9 c/n_p $; $n_s = n(\omega_p/2) = 1.9$, $v_g^s = 0.96 v_g^p$; $n_i = n(\omega_p/2) = 1.8$ $v_g^i = 0.98v_g^p$.
Quasi-phase-matching is obtained with $k_{QPM} = \frac{\omega_p}{2c} (2 n_p - n_s - n_i)$.

Usually, the internal waveguide losses for the TE- and the TM-modes are different~\cite{Melati_2014}. 
In order to take this into account, we introduce two frequency-independent loss coefficients for the signal and idler fields $\eta_s$ and $ \eta_i$, respectively.  
We parametrize these coefficients as  
$
    \bar\eta = \frac{\eta_s+\eta_i}{2},  \, r_\eta = \frac{\eta_s-\eta_i}{\eta_s+\eta_i}.
$
In this paper, we consider three waveguides:
\begin{itemize}
    \item lossless waveguide WG0 with $\eta_s=\eta_i=0$;
     \item waveguide WG1 with equal losses $\eta_s=\eta_i$, $r_\eta=0$;
     \item waveguide WG2 with unbalanced losses for the signal and idler modes $\eta_s=2\eta_i$, $r_\eta=\frac{1}{3}$.
\end{itemize}

In the limit $\bar\eta \rightarrow 0$, both waveguides WG1 and WG2 coincide with the lossless waveguide WG0. We numerically solve the master equations (see Appendix~\ref{appendix_b}, and Ref.~\onlinecite{kopylov_2025_Zenodo})
for each waveguide for different parametric gains $\Gamma$ and losses $\bar\eta$, $r_\eta$. For the resulting correlation matrices $\mathcal{D}(\Gamma,\bar\eta,r_\eta)$ and $\mathcal{C}(\Gamma,\bar\eta,r_\eta)$, we find the Mercer-Wolf, Williamson-Euler, and MSq modes. 
For each basis, we calculate the TMBS with the algorithms described above.

\begin{figure}
    \includegraphics[width=0.32\linewidth]{ 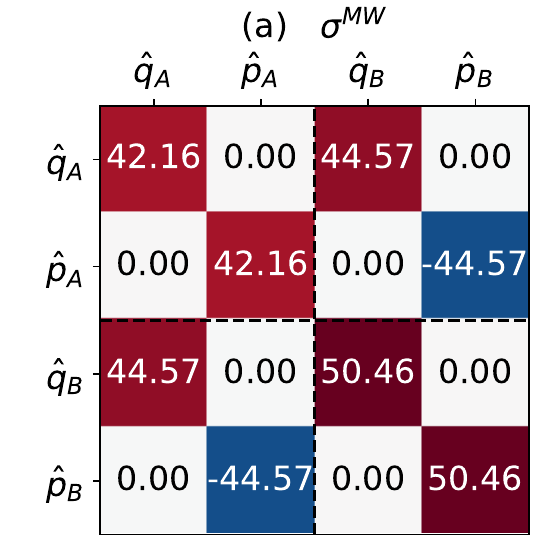 } 
    \includegraphics[width=0.32\linewidth]{ 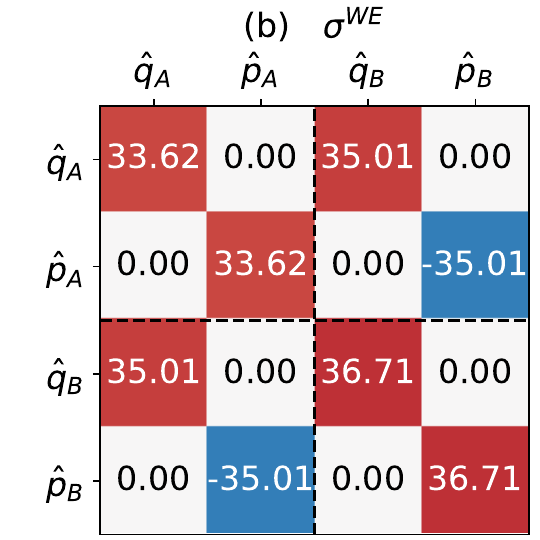 } 
    \includegraphics[width=0.32\linewidth]{ 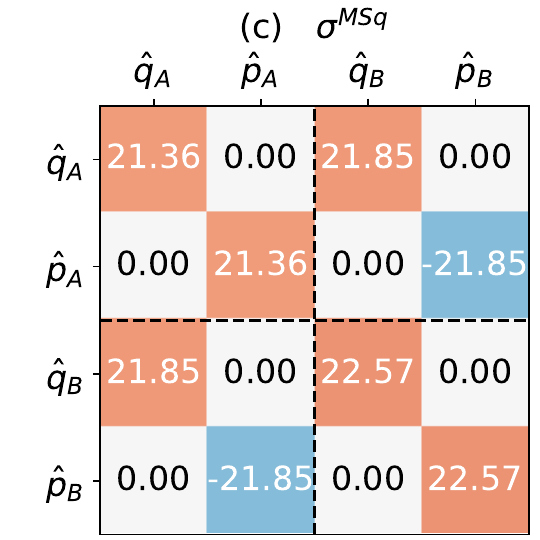 } \\
    \includegraphics[width=0.99\linewidth]{ 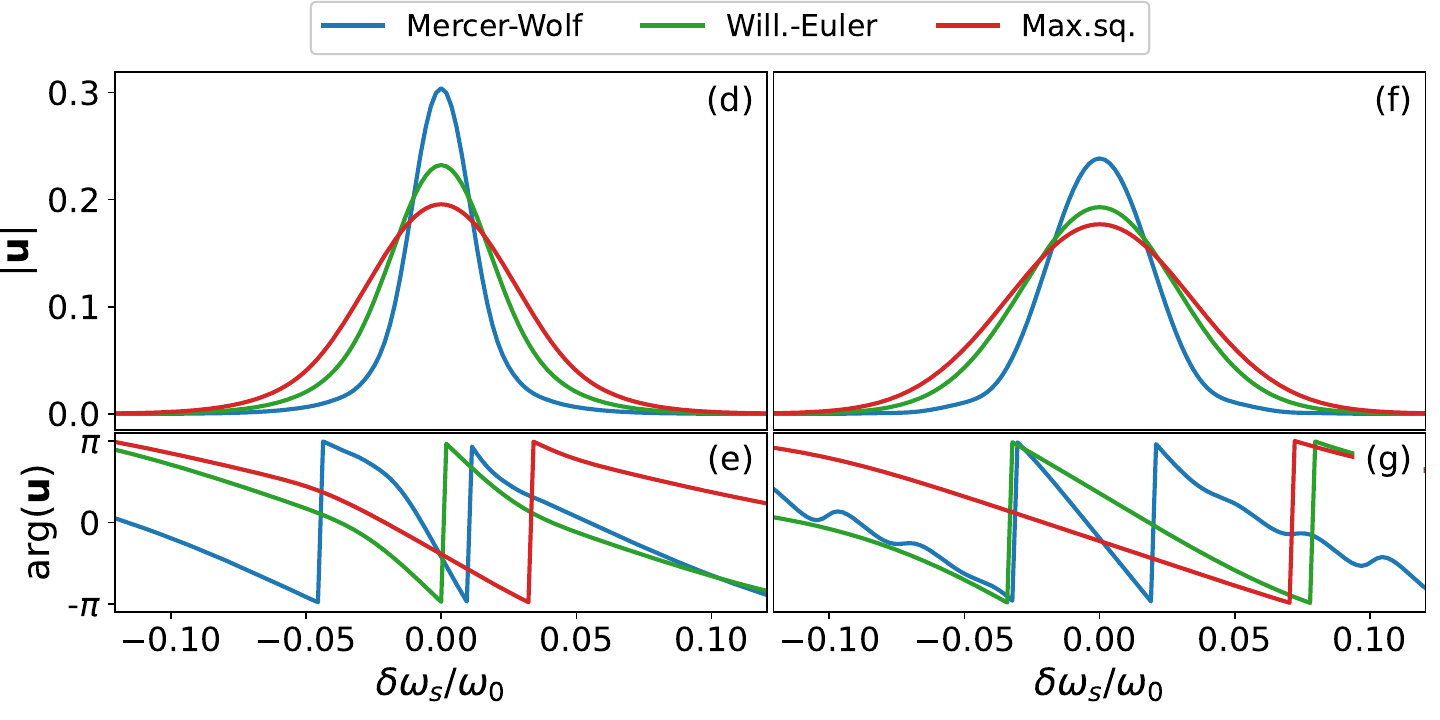 } 
    \caption{ 
        TMBS for the lossy waveguide WG2 (unbalanced losses) with $\bar\eta =5$~dB and $r=\frac{1}{3}$.
        (a,b,c) Covariance matrices $\sigma^{MW}$, $\sigma^{WE}$ and $\sigma^{MSq}$, respectively.
        (d) Absolute value and (e) phase of signal modes $\mathbf{u}_A$;
        (f) absolute value and (g) phase of idler modes $\mathbf{u}_B$. 
        The blue, green, and red colors correspond to the MW-, WE-, and MSq-TMBS bases, respectively.
        Note that the phases of modes are well defined, providing the covariance matrices (a,b,c) to be in the form~Eq.~\eqref{eq_cov_matrix_1x1}.
    }
    \label{fig_4}
\end{figure}

Fig.~\ref{fig_2_lossless} shows results of numerical simulations for the lossless waveguide WG0, for which,
as was mentioned above, there exists the optimal TMBS based on the first Schmidt mode.
Note that the shape of the joint spectral intensity is similar to the results presented for the LiNbO$_3$ based waveguides~\cite{Ebers_2022}.
Fig.~\ref{fig_2_lossless}(b) demonstrates the dependence of the smallest eigenvalue $\lambda_-(\sigma^S)$ (characterized by quadrature variance $\braket{(\Delta \hat{p}^F)^2}$) as a function of the number of photons.
The  obtained TMBS is pure and corresponds to the two-mode squeezed state; therefore, entanglement and squeezing are always present in this state.

In Fig.~\ref{fig_3_alpha_dep}, results of numerical simulations for the lossy waveguides WG1 and WG2 are presented.
The simulations were performed for a fixed gain $\Gamma$, which for the corresponding lossless waveguide WG0 gives the number of photons $N_A=N_B=40$ and the logarithmic negativity $\mathcal{E}(\sigma^S)=5.1$.
In contrast to the lossless case, here the MW-, WE-, and MSq-TMBS provide different quantum states.
In Fig.~\ref{fig_3_alpha_dep}(a, b) the number of photons for the signal and idler parts for different TMBSs are calculated as a function of internal losses $\bar\eta$.
As can be seen, with increasing $\bar\eta$ the number of photons in each TMBS decreases. 
For the waveguides with equal losses, the number of photons in the signal and idler subsystems coincides, while the unbalanced losses lead to different numbers of signal and idler photons in each TMBS. 
Note that the MW-TMBS contains the maximally possible number of photons in each partition.
Figs.~\ref{fig_3_alpha_dep}(c) and (d) demonstrate the dependence of the lowest eigenvalues $\lambda_-(\sigma^{MW})$,  $\lambda_-(\sigma^{WE})$, and  $\lambda_-(\sigma^{MSq})$ on the internal loss coefficient $\bar\eta$ for WG1 and WG2, respectively. 
As expected, the increase of the internal losses $\bar\eta$ decreases both the squeezing and entanglement of all studied TMBS.
However, the maximally entangled state is clearly realized for the MSq-TMBS.
Moreover, the purity of MSq-TMBS is the highest compared to $\sigma^{MW}$ and $\sigma^{WE}$ as is demonstrated in Fig.~\ref{fig_3_alpha_dep}(e,f).

In the case of unbalanced losses, choosing a MW-basis may even lead to a loss of entanglement, as it is demonstrated in Fig.~\ref{fig_3_alpha_dep}(d) for WG2. 
Here, for $\bar\eta > 2.6$~dB, the lowest eigenvalue is grater than unity, $\lambda_-(\sigma^{MW}) >1$, which means that MW-TMBS becomes separable (see the gray region in Fig.~\ref{fig_3_alpha_dep}(d)). For the losses that exhibit the maximum difference in the smallest eigenvalue between the three bases in Fig.~\ref{fig_3_alpha_dep}(d), namely, $\bar\eta = 5$~dB, we present the explicit form of the covariance matrices  $\sigma^{MW}$, $\sigma^{WE}$, and $\sigma^{MSq}$, see Fig.~\ref{fig_4}(a-c), as well as the spectral and phase profiles of $\mathbf{u}_A$ and $\mathbf{u}_B$ modes, see Fig.~\ref{fig_4}(d-g). 
It is clearly seen that the covariance matrix and the spectral width of the modes strongly depend on the chosen basis. 
For example, in the MW-basis, the modes $\mathbf{u}_A$ and $\mathbf{u}_B$ have almost the same width, while in the MSq-basis their widths differ significantly.
The mean number of photons, the lowest eigenvalue, as well as the negativity and purity for this special loss point are presented in Table~\ref{table_1} and show the importance of choosing the correct basis for a particular lossy system.

\begin{table}[h!]
\begin{tabular}{c|ccc}
TMBS & Mer.-Wolf   & Will.-Euler  & MSq   \\
\hline 
$N_A$ & 20.6 & 16.3 & 10.2 \\
$N_B$ & 24.7 & 17.9 & 10.8 \\
\hline
 $\lambda_-(\sigma) = \nu_-(\tilde{\sigma})$  & 1.55  & 0.12  & 0.10 \\
$\braket{(\Delta \hat{p})^2}$, dB & 1.90 & -9.07  & -9.90 \\
$\mathcal{E}(\sigma)$    & 0  & 2.09  & 2.28 \\
\hline
Purity & 0.007  & 0.115  & 0.223 \\
\hline
\end{tabular}
\caption{  The results for different TMBS for the lossy waveguide WG2 (unbalanced losses) with $\bar\eta =5$~dB and $r=\frac{1}{3}$.
} 
\label{table_1}
\end{table}

\section{Summary}

Summing up, we demonstrate how to create the TMBS with the maximal degree of entanglement from a bipartite multimode Gaussian mixed state generated via pulsed type-II multimode PDC in lossy waveguides. 
We show that, in the obtained TMBS, the Gaussian entanglement is characterized by the degree of squeezing and show how the maximally-entangled TMBS can be constructed with the use of the MSq-basis.
We compare this optimal TMBS with the TMBS obtained via the Mercer-Wolf and the Williamson-Euler modes.
By numerical simulations of lossy waveguides, we demonstrate that neither the first Mercer-Wolf mode nor the first Williamson-Euler mode are optimal for preparing the maximally entangled state.

From an experimental point of view, there are three important corollaries.
First, to get the maximally entangled TMBS from the type-II PDC, the measurement basis should coincide with the first mode of the MSq-basis. 
However, this mode does not provide the maximal number of photons, therefore, the intuitive experimental approach of maximizing intensity would not result in a maximally entangled TMBS.
Second, for TMBS, we can use the two-mode squeezing as an experimental measure of entanglement.
Therefore, the proper measurement basis can be found via a diagonalization of the measured covariance matrix.
Alternatively,  one can experimentally implement a multiparameter optimization procedure via Eq.~\eqref{eq_optimization} using the two-mode squeezing degree $\lambda_-\left(\sigma(\mathbf{u}_A, \mathbf{u}_B)\right)$ as a target parameter; however, this procedure requires further investigation, which is out of scope of the present paper.
Third, we performed our numerical simulations only for internal PDC losses, assuming the absence of the external (transmission and detection) losses.
However, as was mentioned above, frequency-dependent external losses can easily be taken into account, 
in particular, when the PDC light passes through spectral filters.
All the procedures and arguments for building an optimal TMBS are also valid for the case of external losses and we leave its the detailed study for the future.

\begin{acknowledgments}
    This work is supported by the `Photonic Quantum Computing' (PhoQC) project, which is funded by the Ministry for Culture and Science of the State of North-Rhine Westphalia. We also acknowledge financial support of the Deutsche Forschungsgemeinschaft (DFG) via the TRR 142/3 (Project No. 231447078, Subproject No. C10).
    We acknowledge support for the publication cost by the Open Access Publication Fund of Paderborn University.
\end{acknowledgments}

\section*{Data Availability Statement}

  Raw data were generated via the numerical solver Ref.~\onlinecite{kopylov_2025_Zenodo}. 

\appendix

\section{ Multimode bipartite Gaussian state }
\label{appendix_a}

A detailed descriptions can be found in Refs.~\onlinecite{Kopylov_2025, Kopylov_2025_PRR}.
Let us consider the multimode Gaussian system, consisting of two parts:  the part A, which is consists of $N$ bosonic modes, denoted as $\hat{\mathbf{a}} = \big(\hat{a}_0, \hat{a}_1, \dots, \hat{a}_N\big)^T$ and the part B, consisting of $M$ bosonic modes $\hat{\mathbf{b}} = \big(\hat{b}_0, \hat{b}_1, \dots, \hat{b}_M\big)^T$.
Commutation relations are bosonic $[\hat{a}_i(z), \hat{a}^\dagger_j(z)]=[\hat{b}_i(z), \hat{b}^\dagger_j(z)]=\delta_{ij} $ and $[\hat{a}_i(z), \hat{b}^\dagger_j(z)]= 0$.
The correlation matrices that fully describe the state of the full system read
\begin{equation}
    \mathcal{D}_{ab} =     
    \begin{pmatrix}
        \braket{\mathbf{\hat{a}^\dagger \hat{a}} }   & \braket{\mathbf{\hat{a}^\dagger \hat{b}} }   \\
        \braket{\mathbf{\hat{b}^\dagger \hat{a}} }   & \braket{\mathbf{\hat{b}^\dagger \hat{b}} }  
    \end{pmatrix}, ~ 
    \mathcal{C}_{ab} =     
    \begin{pmatrix}
        \braket{\mathbf{\hat{a} \hat{a}} }   & \braket{\mathbf{\hat{a} \hat{b}} }   \\
        \braket{\mathbf{\hat{b} \hat{a}} }   & \braket{\mathbf{\hat{b} \hat{b}} }  
    \end{pmatrix}.
    \label{eq_correlation_matrices_DC}
\end{equation}
The expressions in the form $\braket{\mathbf{\hat{a}^\dagger \hat{b}} }$ and $\braket{\mathbf{\hat{a} \hat{b}} }$ denote the $N \times M$ matrices with matrix elements $\braket{\hat{a }_i^\dagger(z) \hat{b}_j(z)}$ and $\braket{\hat{a}_i (z) \hat{b}_j(z)}$, respectively. 

Alternatevely, the state of the full system can be described by the covariance matrix $\Sigma$ with matrix elements
\begin{equation}
  \Sigma_{ij} = \dfrac{\braket{\hat{x}_i\hat{x}_j + \hat{x}_j\hat{x}_i }}{2}-\braket{\hat{x}_i}\braket{\hat{x}_j},
\end{equation}
where the vector $\hat{\mathbf{x}}  = (\hat{q}^a_1, \hat{p}^a_1, \dots, \hat{q}^a_N, \hat{p}^a_N; \hat{q}^b_1, \hat{p}^b_1, \dots, \hat{q}^b_M, \hat{p}^b_M)^T $, while the quadrature operators are $\hat{q}^c_i = \hat{c}^\dagger_i + \hat{c}_i $ and $\hat{p}^c_i =  i(\hat{c}^\dagger_i - \hat{c}_i) $ with the commutation relations $[\hat{q}^c_n, \hat{p}^d_m] = 2i\delta_{nm}\delta_{cd}$. 
Here, the upper indexes indicate arbitrary modes $c,d =(a, b)$.
The elements of the covariance matrix are connected with the correlation matrices~\eqref{eq_correlation_matrices_DC} as 
 \begin{align}
      \label{eq_cov_matrix_determination_1}
    \braket{ \hat{q}^c_i \hat{q}^d_j} = 
    \delta_{ij}\delta_{cd} 
     + 2 \Big(\text{Re}[\braket{\hat{c}^\dagger_i \hat{d}_j}]
     + \text{Re}[\braket{\hat{c}_i \hat{d}_j}] \Big), \\
          \label{eq_cov_matrix_determination_2}
    \braket{ \hat{p}^c_i \hat{p}^d_j}  = 
    \delta_{ij}\delta_{cd} 
     + 2 \Big(\text{Re}[\braket{\hat{c}^\dagger_i \hat{d}_j}]
     - \text{Re}[\braket{\hat{c}_i \hat{d}_j}] \Big), \\
       \label{eq_cov_matrix_determination_3}
    \dfrac{ \braket{ \hat{p}^c_i \hat{q}^d_j +  \hat{q}^d_j \hat{p}^c_i} }{2} =  2 \Big(\text{Im}[\braket{\hat{c}_i \hat{d}_j}] -\text{Im}[\braket{\hat{c}^\dagger_i \hat{d}_j}]   
       \Big), \\
       \braket{\hat{q}^c_i}=2\text{Re}[\braket{\hat{c}_i}], ~ ~ ~
       \braket{\hat{p}^c_j}=2\text{Im}[\braket{\hat{c}_j}].
\end{align}

The general local passive unitary transformation is defined as $U = U_A \oplus U_B$, where the new broadband modes are  $\hat{\mathbf{A}} = U_A \hat{\mathbf{a}}$ and $\hat{\mathbf{B}} = U_B \hat{\mathbf{b}}$.
The correlation matrices in the new broadband modes are
\begin{equation}
    \mathcal{D}_{AB} = U^* \mathcal{D}_{ab} U^T, \text{ ~  and  ~ } \mathcal{C}_{AB} = U \mathcal{C}_{ab} U^T.
    \label{eq_general_transformation}
\end{equation}

In addition to the creation and annihilation operators, we also need the  quadrature operators for new broadband modes.
For an arbitrary mode with the annihilation operator $\hat{F}_n$, the quadrature operators are $\hat{q}^F_n = \hat{F}^\dagger_n + \hat{F}_n $ and $\hat{p}^F_n =  i(\hat{F}^\dagger_n - \hat{F}_n) $ with the commutation relations $[\hat{q}^F_n, \hat{p}^F_m] = 2i\delta_{nm}$.

\section{ Master equation }
\label{appendix_b}

The type-II PDC in a lossy waveguide can be described in terms of a master equation of the correlation matrices~\eqref{eq_correlation_matrices_DC}.
The signal and idler fields are taken in a discrete uniform frequency space $(\omega_0, \omega_1, \dots, \omega_N)$, which allows us to describe signal and idler systems by $N$ monochromatic modes $\hat{\mathbf{a}}(z)$ and $N$ monochromatic modes $\hat{\mathbf{b}}(z)$, respectively.
The spatial master equation for PDC correlation matrices has the form 
\begin{multline}
    \dfrac{d \mathcal{D}(z)}{d z} =  i \big(\mathcal{D}(z) K - K^* \mathcal{D}(z)\big)  \\ + i \Gamma \big(\mathcal{C}^{*}(z) M^T(z) -  M^*(z) \mathcal{C}(z) \big),
    \label{eq_master_equation_1}
\end{multline}
\begin{multline}
    \dfrac{d \mathcal{C}(z) }{d z} = i \big(\mathcal{C}(z) K + K \mathcal{C}(z)\big) \\ + i\Gamma   \Big( \big(M(z) \mathcal{D}(z)+M(z) \big)^T + M(z) \mathcal{D}(z)  \Big),
    \label{eq_master_equation_2}
\end{multline}
where the superscript $[\cdot]^*$ denotes the complex conjugation of a matrix. 
The matrix $K$ is a diagonal matrix consisting of  complex wave vectors  $\textrm{diag}(\kappa^a_0, \dots \kappa^a_N, \kappa^b_0, \dots, \kappa^b_N)$. Here, $\kappa^a_n = k^a_n+i\eta_a/2$, where $k^a_n$ is a real wave vector for mode $\hat{a}_n$ and $\eta_a$ is the loss coefficient for subsystem $a$. 
The $z$-dependent matrix $M(z)$ is given by
\begin{equation}
    M(z)=
    \begin{pmatrix}
        \mathbf{0}_{N}       & J(z)  \\
        J^{T}(z)  &   \mathbf{0}_{N} 
    \end{pmatrix}
    \label{eq_coupling_matrix}
\end{equation}
 and is a coupling matrix of a spatial generator for the PDC process $\hat{G}_{pdc}(z) =   \frac{1}{2}\hbar\Gamma \sum_{i,j} J_{ij}(z)  \hat{a}_i^\dagger(z) \hat{b}_j^\dagger(z)  + h.c.
$
Explicitly, the matrix $J_{ij}(z) = S(\omega_i + \omega_j) e^{i (k_p(\omega_i + \omega_j) - k_{QPM})z } $; $S(\omega)$ is the pump spectral distribution at $z=0$; $k^{(a,b,p)}_n \equiv k^{(a,b,p)}(\omega_n) = \frac{n^{(a,b,p)}(\omega_n)\omega_n}{c}$ are the wave vectors of the (a) signal, (b) idler, and (p) pump fields in the waveguide; $k_{QPM} = 2\pi/\Lambda$ and $\Lambda$ is the poling period for the quasi-phase-matching condition. 
The parameter $\Gamma$ determines the coupling strength of the PDC process.

The initial condition (vacuum state) reads $\mathcal{D}(0)=\mathcal{C}(0)=\mathbf{0}_{2N}$ and,
together with the coupling matrix in the form of Eq.~\eqref{eq_coupling_matrix}, determines the structure of the solution in the form of Eq.~\eqref{eq_solution_block}.

\section{ Mercer-Wolf and Williamson-Euler decompositions }
\label{appendix_c}

\paragraph{Mercer-Wolf modes}

can be obtained via the diagonalization of the matrix $\mathcal{D}$~\cite{Kopylov_2025, Kopylov_2025_PRR}.
Its diagonalization can be obtained via the independent eigendecomposition of the matrices $\braket{\mathbf{\hat{a}^\dagger \hat{a}} } = V_A \Lambda_A V_A^H$ and $\braket{\mathbf{\hat{b}^\dagger \hat{b}} }  = V_B \Lambda_B V_B^H $, where the diagonal matrices $\Lambda_A$ and $\Lambda_B$ are given in decreasing order.

The TMBS, based on the Mercer-Wolf modes (MW-TMBS) is built for the vectors $\mathbf{u}^{MW}_A =  e^{i\frac{\phi}{2}} \mathbf{v}_A$ and $\mathbf{u}^{MW}_B =  e^{i\frac{\phi}{2}} \mathbf{v}_B$, where the row-vectors $\mathbf{v}_A$ and $\mathbf{v}_B$  are the first rows of the matrices $V_A^T$ and $V_B^T$, respectively, and a constant phase $\phi = \mathrm{arg} \big( \mathbf{v}_A \braket{\mathbf{\hat{a} \hat{b}} } \mathbf{v}_B^T \big)$.
The covariance matrix of MW-TMBS is defined as $\sigma^{MW} \equiv \sigma(\mathbf{u}^{MW}_A, \mathbf{u}^{MW}_B)$.
As long as the modes are obtained via the diagonalization of matrix $\mathcal{D}$, the first Mercer-Wolf modes $\mathbf{u}^{MW}_A$ and $\mathbf{u}^{MW}_B$ contain the maximally possible number of photons per mode.

\paragraph{Williamson-Euler modes}

are obtained via the  Williamson-Euler decomposition~\cite{Houde_2023,Kopylov_2025}, namely a decomposition of the initial covariance matrix $\sigma = O_l \Lambda O_r  D O_r^T \Lambda  O_l^T$, where the matrices
$O_l$ and $O_r$ are orthogonal and matrices  $D=\text{diag}(\nu_1, \nu_1, \dots, \nu_{N}, \nu_{N})$ and $\Lambda=\text{diag}(e^{r_1}, e^{-r_1}, \dots, e^{r_{N}},e^{-r_{N}})$ are diagonal; the values $r_i$ are sorted in descending order.
Having the first vector $\mathbf{v}^{(1)} = (x^{a}_1, y^{a}_1, \dots, x^{a}_N, y^{a}_N; x^{b}_1, y^{b}_1, \dots, x^{b}_M, y^{b}_M )$ of matrix $O_l$,
and applying the same procedure as for the MSq mode, we can build modes for each part as $ \hat{A}^{WE} = \mathbf{u}^{WE}_A \mathbf{\hat{a}}  $ and $ \hat{B}^{WE} = \mathbf{u}^{WE}_B \mathbf{\hat{b}} $, 
where $\mathbf{u}^{WE}_A = \frac{\mathbf{v}^a}{|\mathbf{v}^a|} $ and  $\mathbf{u}^{WE}_B =  \frac{\mathbf{v}^b}{|\mathbf{v}^b|} $ with $[\mathbf{v}^a]_n = y^a_n+i x^a_n$ and $[\mathbf{v}^b]_n = y^b_n+i x^b_n$. 
Numerically, the Williamson-Euler decomposition can be found with the use of algorithms presented in, e.g., Refs.~\cite{Safranek_2018,Gupt2019,houde_2024}.
The TMBS obtained by the vectors $\mathbf{u}^{WE}_A$ and $\mathbf{u}^{WE}_B$ is further called as WE-TMBS;  the corresponding covariance  matrix is defined  as $\sigma^{WE} \equiv \sigma(\mathbf{u}^{WE}_A, \mathbf{u}^{WE}_B)$.

\bibliography{lit_gauss_entanglement.bib}

\end{document}